\documentclass{aa}

\usepackage{txfonts}
\usepackage{amsmath}
\usepackage{amssymb,amsfonts,textcomp}
\usepackage{threeparttable}
\usepackage{afterpage}
\usepackage{savesym}
\savesymbol{tablenum}
\usepackage{siunitx}
\restoresymbol{SIX}{tablenum}

\usepackage{algorithmic}
\usepackage{algorithm}
\floatname{algorithm}{Algorithm}
\usepackage{varwidth}
\usepackage{makecell}

\newcommand{\beq}{\begin{equation}}
\newcommand{\eeq}{\end{equation}}
\newcommand{\beqn}{\begin{eqnarray}}
\newcommand{\eeqn}{\end{eqnarray}}
\DeclareMathOperator*{\argmin}{arg\,min}

\DeclareSIUnit \Jansky {Jy}
\usepackage[normalem]{ulem}

\begin{document}

\title{Direction-dependent calibration with image-domain gridding}
\titlerunning{Calibration with IDG}

\author{Sebastiaan van der Tol\inst{1}
\and Sarod Yatawatta\inst{1}
\and Bram Veenboer\inst{1}
\and David Rafferty\inst{2}}

\institute{
  Netherlands Institute for Radio Astronomy (ASTRON), Postbus 2, 7990 AA Dwingeloo, The Netherlands
  \and University of Hamburg, Gojenbergsweg 112, 21029 Hamburg, Germany}

\abstract
  {}
  {
    Wide-field images made by radio interferometers are invariably affected by direction-dependent systematic effects such as the ionosphere or the beam pattern. Calibration along a set of discrete directions in the sky is the default technique to estimate and correct these systematic errors. However, additional processing such as smoothing and mosaicing are required to reconcile the step wise variation of the estimated systematic errors at the edges of the discrete directions (facets).
  }
  {
We overcome the discrete nature of direction-dependent calibration by using image-domain gridding as the model for the calibration. Instead of discrete directions in the sky, calibration is performed using a basis that represents a set of sub-grids in the Fourier space. This automatically removes the need for extra operations to recover the wide-field systematics error model without any discontinuity.
  }
  {
    We provide results based on LOFAR data where we compare the traditional facet-based (discrete directional gains) calibration with the proposed approach. The comparison shows improved image quality, mainly because of the physical plausibility of the proposed approach as opposed to using a piecewise constant model for direction-dependent systematic errors.
  }
  {}
   \keywords{
     Instrumentation: interferometers --
   Techniques: interferometric
               }

\maketitle

\section{Introduction}
Radio interferometers that have wide fields of view, for example the low frequency array \citep[LOFAR;][]{LOFAR} and the Murchison wide-field array \citep[MWA;][]{MWA}, require specialized data processing for the estimation and correction of direction-dependent systematic errors across their observing fields. Direction-dependent calibration is an essential step in this data-processing chain, and there are a plethora of methods that accomplish this such as \cite{Kaz2}, \cite{DCAL}, \cite{vanWeeren2016}, and \cite{Tasse2018}. The application of corrections for these systematic errors during image synthesis is an equally important step in the data processing chain, and there are also many wide-field imaging and deconvolution methods that accomplish this, including \cite{WBA}, \cite{B2008}, \cite{Aproj}, \cite{WSClean}, and \cite{IDG}.

While there are advanced methods for (i) direction-dependent calibration and for (ii) wide-field imaging with correction for direction-dependent errors, they are mostly disjointed. In other words, methods and software that fall into either of these categories have been developed with a focus exclusively on direction-dependent calibration or on wide-field imaging with correction for direction-dependent errors. A typical image synthesis from raw data is performed by first running direction-dependent calibration (estimating systematic errors) and thereafter running image synthesis (correcting for systematic errors in the data). Direction-dependent calibration is performed along discrete directions in the sky, with the estimated systematic errors only being applicable to a small area in the sky centered around this direction. This area is normally called a facet. After making images of each facet using image-synthesis techniques that incorporate direction-dependent effects, such facets are combined using techniques such as mosaicing \citep{Sault1996} to build the full field-of-view image. With minor deviations, the aforementioned scheme is the core data-processing strategy currently used by LOFAR imaging \citep{vanWeeren2016,Tasse2018}.

The unification of direction-dependent calibration and imaging has seen limited growth compared to the disjointed methods that specialize in either calibration or imaging. This is mainly due to the complexity of joint calibration and imaging in a computationally feasible manner. Most joint methods in calibration with imaging parameterize both the sky and the instrument systematics and find a joint solution.  Due to the restriction in computational complexity, existing methods that fall into the category of joint direction-dependent calibration and imaging are few in number, can only handle simple physical models for the systematics, or can only handle narrow fields of view \citep[e.g.,][]{Intema,Arras2019,Birdi2020,Roth2023}. In this work, we overcome most of these limitations by proposing the use of image-domain gridding \citep[IDG;][]{IDG} as the forward model prediction in direction-dependent calibration. In comparison to existing methods for joint calibration and imaging, the proposed method is different because we still have separate calibration (instrument-parameterized) and imaging (sky-parameterized) steps. During the calibration step, we estimated the direction-dependent errors covering the full field of view, parameterized using an arbitrary basis in space. The estimated A-terms (or systematic error terms) are thereafter directly applied during the image synthesis, without having an intermediate systematic-error model creation step. In order to reduce the computational complexity in calibration, we use stochastic optimization methods \citep{DSW2019} that can minimize optimization problems with a large parameter space with super-linear convergence. A noteworthy similarity of the proposed method partially exists with the method by \cite{Roth2023} in the use of IDG for forward model prediction. However, the proposed method is computationally more efficient than that of \cite{Roth2023} because we follow a frequentist approach (maximum likelihood) rather than a Bayesian approach, and therefore our method is compatible with wide fields of view.

The remainder of this paper is organized as follows. In Section \ref{sec:model} we give an overview of IDG and how it is used in calibration (IDG-CAL). Next, in Section \ref{sec:calibration} we describe the stochastic optimization scheme used in calibration. In Section \ref{sec:results} we provide results based on LOFAR observations to illustrate the efficacy of IDG-CAL compared to the conventional facet-based calibration. We discuss the computational and physical advantages of the proposed method compared to alternatives in Section \ref{sec:discussion}. Finally, we draw our conclusions in Section \ref{sec:conclusions}.

Please note that lowercase bold letters refer to column vectors (e.g., ${\bf y}$). Uppercase bold letters refer to matrices (e.g., ${\bf C}$). 
Indices are lower case letters. Where possible, the number of elements
over which an index runs is indicated by the same letter in uppercase.
Unless otherwise stated, all parameters are complex numbers. The set of complex numbers is given as ${\mathbb C,}$ and the set of real numbers is given as  ${\mathbb R}$. The imaginary unit is denoted by $\jmath$ (a dotless $j$). The matrix Hermitian transpose is referred to as $(\cdot)^{H}$. The matrix Kronecker product is given by $\otimes,$ and the matrix Hadamard product is given by $\odot$. The identity matrix of size $N$ is given by ${\bf I}_N$. The Frobenius norm is given by $\|\cdot \|$.
The function $\operatorname{vec}(\cdot)$ stacks the columns of a matrix in a vector, and $\operatorname{diag}(\cdot)$ makes a diagonal matrix from a vector.

\section{Data model\label{sec:model}}
In this section, we provide a brief introduction to IDG \citep{IDG,Veenboer2017} and how it is used as the forward model prediction for direction-dependent calibration.
\subsection{Image-domain gridding}
Image domain gridding \citep{IDG} computes the equivalent of A-projection \citep{B2008} while avoiding the computation of over-sampled convolution kernels. This is advantageous in situations where many different kernels are needed for relatively few
visibilities per kernel.

With regard to correction for direction-dependent errors in image synthesis, both A-projection and IDG offer the advantage of applying a continuous correction over facet-based methods that apply piecewise constant corrections. However, since traditional direction-dependent calibration methods are facet-based, and interpolation from facet-based corrections to an A-term introduces new errors, the A-projection and related methods have (so far) mostly been used to apply analytical beam models only, for which the A-term can be readily computed.

The calibration method IDG-CAL presented in this paper is based on the realization that IDG is sufficiently efficient in applying varying A-terms and that it
is feasible to use IDG iteratively in a calibration scheme to compute derivatives and residuals.
This allows us to solve for A-terms directly from the visibilities, instead of from intermediate solutions.

Per baseline, the visibilities are partitioned in time-frequency ranges, $T_p$, such that
the uv-coverage convolved by a kernel fits within sub-grid $p$.
The kernel includes a taper, the A-term, and a differential w-term.
A sub-grid consists of $Q=N_g\times N_g$ pixels (typically 32$\times$32). 
For efficiency reasons, $N_g$ is always chosen to be a multiple of eight.
Each sub-grid is extracted from the Fourier transform of the full model image 
and then transformed back to the image domain. 
A sub-grid belongs to a single baseline. Per baseline, there is at least one sub-grid, but there can be more.
Sub-grids can overlap and be placed arbitrarily on the uv (or Fourier space) grid.

Once the sub-grids are known, only a single equation remains to compute the model visibilities. The calibration algorithm presented in this paper was 
derived from the equation provided below.
The model visibilities, $\mathbf{V}_{jk}$ ($\in \mathbb{C}^{2\times 2}$), for baseline $j$ and time-frequency index $k$ are given as a sum over the pixels of sub-grid $p$ as
\begin{equation}
  \mathbf{V}_{jk} = \sum_{q=1}^{Q} \mathbf{A}_{s_{1p}q} \mathbf V_{pq} \mathbf{A}_{s_{2p}q}^{\mathrm{H}}\varphi_{pqk}
  \label{eq:model_visibilities}
,\end{equation}
where ${s_{1p}}$ and ${s_{2p}}$ are the station (or antenna) indices (total $S$ stations) corresponding to baseline $j_p$ associated with sub-grid $p$. In the context of LOFAR data processing, a station is essentially an electronically steerable antenna.
The value of pixel $q$ of sub-grid $p$ in the image domain is denoted by $\mathbf V_{pq}$. The A-terms representing the systematic errors of stations $s_{1p}$ and $s_{2p}$ on sub-grid $p$, evaluated at pixel $q,$ are given by $\mathbf{A}_{s_{1p}q}$ and $\mathbf{A}_{s_{2p}q}$, respectively. Note that $\mathbf{A}_{s_{1p}q}$,$\mathbf{A}_{s_{2p}q}$, $\mathbf{V}_{pq}$ $\in\mathbb{C}^{2\times 2}$. The complex exponential phase term is given by $\varphi_{pqk}$. This term represents the propagation delay, which depends on the sub-grid pixel position $(l_q, m_q)$ and the $u_{pk}, v_{pk}, w_{pk}$ coordinates of the visibility $\mathbf{V}_{jk}$.

Note that form (\ref{eq:model_visibilities}) is exactly the same as the faceted approach. The only difference here is that the summation is over sub-grid pixels and not facets.
Effectively, IDG applies a correction that is interpolated from sub-grid resolution to full image resolution.
This approaches physical reality much closer than what can be obtained using a facet-based approach.

\subsection{Parameterization of A-terms}
An A-term in IDG is sampled at sub-grid resolution.
To limit the number of degrees of freedom, the A-terms are written as an expression of a few parameters using an appropriate basis. The parameterization is rather arbitrary, since IDG-CAL does not depend on the precise choice of the basis as long as the derivatives with respect to the parameters can be computed.

A simple example of a parameterization (a scalar A-term) can be given as
\begin{equation} \label{Apoly}
  \mathbf{A}_q(\mathbf{x}) = \left(\sum_{r=1}^{R_a} a_r h_r(l_q,m_q) \right) \operatorname{exp}\left(\jmath \sum_{r=1}^{R_b} b_r h_r(l_q,m_q)) \right) \mathbf{I_2},
\end{equation}
where $\jmath$ denotes the imaginary unit ($\jmath^2=-1$). The polynomial basis is expressed as $h_1(l,m) = 1$, $h_2(l,m) = l$, $h_3(l,m) = m$, $h_4(l,m) = l^2, h_5(l,m) = lm$, $h_6(l,m) = m^2$, and so on.
In (\ref{Apoly}), $q$ is the sub-grid pixel index and $l_q$ and $m_q$ are the image coordinates of that pixel.
The parameters, $\mathbf{x,}$ are partitioned into parts for the amplitude, $\mathbf{a,}$ and for the phase, $\mathbf{b}$:
\begin{equation}
  \mathbf{x} = 
  \left[
    \begin{array}{c}
      \mathbf{a} \\
      \mathbf{b}
    \end{array}
  \right],
\end{equation}
where $\mathbf{a}$ is a vector of length, $R_a$, and $\mathbf{b}$ is a vector of length, $R_b$, with real parameters.
The total number of parameters for all $P$ sub-grids and $S$ stations is therefore $R=(R_a+R_b)\times S$. Note that a more detailed parameterization can be used to model diagonal A-terms or A-terms for full polarization, but we left out this detail for simplicity.

The input data are the vector, $\mathbf{y,}$ containing all observed visibilities that are being calibrated.
The function $\mathbf{\hat{V}}_{jk}({\bf y})$ selects the visibilities corresponding
to the baseline, $j,$ and time-frequency index, $k,$ from input $\mathbf{y}$ and returns a matrix $(\in \mathbb{C}^{2 \times 2})$. Likewise, the model visibilities can be written as function of all the free
parameters as $\mathbf{{V}}_{jk}({\bf x})$. This function only truly depends
on a subset of $\mathbf{x}$, i.e., the entries corresponding to 
baseline $j$ and time frequency index $k$.
The model visibilities also depend on the model image, through the sub-grids.
This dependence is not included in the notation because the model image is not a free parameter that is going to be solved for here in this formulation.
The difference between the observed visibilities and the model visibilities are the residual visibilities given by
\begin{equation}
  \mathbf{\tilde{V}}_{jk}({\bf x},{\bf y})= \mathbf{\hat{V}}_{jk}({\bf y}) - \mathbf{{V}}_{jk}({\bf x}).
  \label{eq:residual_visibilities}
\end{equation}

The goal of the calibration algorithm is to find the parameters that minimize a cost function measuring the error in terms of the norm of the residual.
The cost function to be minimized is designed as a weighted sum of squares of the residual visibilities (under a Gaussian noise model, summed over $J$ baselines and $K$ time-frequency samples):
\begin{equation} \label{cost_function}
  f({\bf x},{\bf y}) = \sum_{j=1}^{J} \sum_{k=1}^{K}
    \operatorname{vec}\left(\mathbf{\tilde{V}}_{jk}({\bf x},{\bf y})\right)^{\mathrm{H}}
    \mathbf{\overline{W}}_{jk}
    \operatorname{vec}\left(\mathbf{\tilde{V}}_{jk}({\bf x},{\bf y})\right)
,\end{equation}
where $\mathbf{\overline{W}}_{jk}=\operatorname{diag}\left(\operatorname{vec}\left(\mathbf{W}_{jk}\right)\right)$ 
contain the data weights (given a priori).

Note that any given sub-grid belongs to a single baseline and a time-frequency selection.
Since there are $P$ sub-grids in total, the cost function can be split into a sum over per-sub-grid cost functions as
\begin{equation} \label{subgrid_cost}
  f({\bf x},{\bf y}) = \sum_{p=1}^{P} C_p({\bf x},{\bf y}).
\end{equation}The derivative of the cost function (\ref{cost_function}) can be calculated in closed form by considering the per-sub-grid cost function (\ref{subgrid_cost}) with respect to parameter $x_{sr}$ for station $s$ and parameter index $r$ as
\begin{equation}
  \frac{\partial}{\partial x_{sr}} f({\bf x},{\bf y}) = \sum_{p \in P_s} \frac{\partial}{\partial x_{sr}} C_p({\bf x},{\bf y}),
\end{equation}
where $P_s$ is a set of indices to sub-grids in which station $s$ participates.

\subsection{Per-sub-grid equations of the derivative\label{ssec:derivative}}
In order to calculate the closed form derivative, we considered the separable, sub-grid-based cost function (\ref{subgrid_cost}). Without loss of generality, we only considered a single sub-grid, $p,$ for simplicity. The associated time-frequency range is denoted by $T_p$, the baseline $j_p$, and the stations $s_{1p}$ and $s_{2p}$.

The cost function associated with sub-grid $p$ is given by
\begin{equation}
  C_p\left({\bf x},{\bf y}\right) = \sum_{k \in T_p}
    \operatorname{vec}\left(\mathbf{\tilde{V}}_{j_pk}({\bf x},{\bf y})\right)^{\mathrm{H}}
    \mathbf{\overline{W}}_{j_pk}
    \operatorname{vec}\left(\mathbf{\tilde{V}}_{j_pk}({\bf x},{\bf y})\right).
\end{equation}The derivative of the per-sub-grid cost function is given by
\begin{align}
  & \frac{\partial}{\partial x_{sr}} C_p\left({\bf x},{\bf y}\right) = \nonumber \\
  & \quad \sum_{k \in T_p} \left(
  \operatorname{vec}\left(\frac{\partial}{\partial x_{sr}}\mathbf{\tilde{V}}_{j_pk}({\bf x},{\bf y})\right)^{\mathrm{H}}
 \mathbf{\overline{W}}_{j_pk}
 \operatorname{vec}\left(\mathbf{\tilde{V}}_{j_pk}({\bf x},{\bf y})\right) + \right. \nonumber \\
 & \quad \quad \left.
 \operatorname{vec}\left(\mathbf{\tilde{V}}_{j_pk}({\bf x},{\bf y})\right)^{\mathrm{H}}
 \mathbf{\overline{W}}_{j_pk}
 \operatorname{vec}\left(\frac{\partial}{\partial x_{sr}}\mathbf{\tilde{V}}_{j_pk}({\bf x},{\bf y})\right) 
 \right) = \nonumber \\
 & \quad 2 \operatorname{Re} \sum_{k \in T_p}
  \operatorname{vec}\left(\frac{\partial}{\partial x_{sr}}\mathbf{\tilde{V}}_{j_pk}({\bf x},{\bf y})\right)^{\mathrm{H}}
 \mathbf{\overline{W}}_{j_pk}
 \operatorname{vec}\left(\mathbf{\tilde{V}}_{j_pk}({\bf x},{\bf y})\right) = \nonumber \\
 & \quad - 2 \operatorname{Re} \sum_{k \in T_p}
  \operatorname{vec}\left(\frac{\partial}{\partial x_{sr}}\mathbf{V}_{j_pk}({\bf x},{\bf y})\right)^{\mathrm{H}}
 \mathbf{\overline{W}}_{j_pk}
 \operatorname{vec}\left(\mathbf{\tilde{V}}_{j_pk}({\bf x},{\bf y})\right).
 \label{eq:derivative_cost_function}
\end{align}

The derivative with respect to $x_{s_{1p}r}$ of the model visibilities  is given by
\begin{equation}
  \frac{\partial}{\partial x_{s_{1p}r}} \mathbf{V}_{j_pk} = \sum_{q=1}^{Q} \frac{\partial}{\partial x_{s_{1p}r}} \mathbf{A}_{s_{1p}q} \mathbf V_{pq} \mathbf{A}_{s_{2p}q}^{\mathrm{H}} \varphi_{pqk}
  \label{eq:derivative_visibilities},
\end{equation}
for all $k \in T_p$. The derivative with respect to the parameters for the other station, $x_{s_{2p}r}$,
is given in similar fashion, but is not worked out here.
In $\operatorname{vec}\left(\cdot\right)^{\mathrm{H}}$ form, this becomes
\begin{equation}
  \frac{\partial}{\partial x_{s_{1p}r}} \operatorname{vec}\left(\mathbf{V}_{jk}\right)^{\mathrm{H}} = 
  \sum_{q=1}^{Q} \operatorname{vec} \left( \mathbf V_{pq}  \right)^{\mathrm{H}} \left(\mathbf{A}_{s_{2p}q}^{*} \otimes \frac{\partial}{\partial x_{s_{1p}r}} \mathbf{A}_{s_{1p}q} \right)^{\mathrm{H}} \varphi_{pqk}^{*}.
  \label{eq:vec_derivative_visibilities}
\end{equation}
Substituting Eq. \eqref{eq:vec_derivative_visibilities} into Eq. \eqref{eq:derivative_cost_function} gives
\begin{align}
  & \frac{\partial}{\partial x_{s_{1p}r}} C_p\left({\bf x},{\bf y}\right) = - 2 \operatorname{Re} \sum_{k \in T_p}
  \sum_{q=1}^{Q} \left(\rule{0cm}{2em}\right. \operatorname{vec} \left( \mathbf V_{pr}  \right)^{\mathrm{H}} \times \nonumber \\
  & \quad  \left(\mathbf{A}_{s_{2p}q}^{*} \otimes \frac{\partial}{\partial x_{s_{1p}r}} \mathbf{A}_{s_{1p}q} \right)^{\mathrm{H}} \varphi_{pqk}^{*}
   \mathbf{\overline{W}}_{j_pk}
   \operatorname{vec}\left(\mathbf{\tilde{V}}_{j_pk}({\bf x},{\bf y})\right) \left.\rule{0cm}{2em}\right).
   \label{eq:derivative_cost_function_expanded}
\end{align}
Then, we define the weighted residual as
\begin{equation}
  \mathbf{X}_{pq} = \sum_{k \in K_p} \varphi_{j_pkq}^{*} \mathbf{W}_{j_pk} \odot \mathbf{\tilde{V}}_{j_pk}({\bf x},{\bf y}).
  \label{eq:accumulated_residual_visibilities}
\end{equation}
Rearranging Eq. \eqref{eq:derivative_cost_function_expanded} and writing it in terms of the accumulated residual visibilities,
$\mathbf{X}_{pq}$, this becomes
\begin{equation}
  \frac{\partial}{\partial x_{s_{1p}r}} C_p\left({\bf x},{\bf y}\right) =
  -2 \operatorname{Re}
    \sum_{q=1}^{Q} \operatorname{vec} \left( \mathbf V_{pq}  \right)^{\mathrm{H}}  
    \operatorname{vec}\left(\frac{\partial}{\partial x_{s_{1p}r}} \mathbf{A}_{s_{1p}q}^{\mathrm{H}}\mathbf{X}_{pq} \mathbf{A}_{s_{2p}q} \right).
  \label{eq:derivative_cost_function_final}    
\end{equation}The derivation above leads to Algorithm \ref{alg_derivatives} for the computation of the derivatives.

\begin{algorithm}
  \caption{Derivative computation}\label{alg_derivatives}
  \begin{algorithmic}[1]
    \REQUIRE Model image, sub-grid size
    \STATE Fourier transform the model image to the uv-domain.
    \STATE Split off sub-grids from uv-grid.
    \STATE Fourier transform sub-grids to the image domain.
    \STATE Apply the known effects (beam, smearing) to the sub-grids.
    \STATE Compute the model visibilities from the sub-grids according to \eqref{eq:model_visibilities} and subtract them from the observed visibilities
    according to \eqref{eq:residual_visibilities}
    to obtain the residual visibilities. \label{step:compute_model_visibilities}
    \STATE Accumulate the residual visibilities onto sub-grids as per \eqref{eq:accumulated_residual_visibilities}.
    \STATE Compute the per-sub-grid contribution to the derivatives per \eqref{eq:derivative_cost_function_final} and accumulate them to obtain the total derivative.
    \end{algorithmic}
\end{algorithm}

\noindent Note that this procedure is analogous to the way the residual image is computed with IDG. This analogy is not surprising since the residual image is the derivative of cost function \eqref{cost_function} with respect to the pixel values of the model image \citep{Rau2009}.

When only the A-term has changed, the updated derivative can be computed by starting from step \ref{step:compute_model_visibilities}. In Section \ref{sec:calibration}, we discuss how the calculated derivative is used to minimize the calibration cost function (\ref{cost_function}).

\section{Stochastic calibration\label{sec:calibration}}
In this section, we describe the optimization strategy used for IDG-CAL for computational efficiency and stability. We used the closed form expressions for the cost function (\ref{cost_function}) as well as its derivative as given in Algorithm \ref{alg_derivatives} for this purpose.
We considered calibration using IDG to be equivalent to minimizing the nonlinear, non-convex cost function $f({\bf x},{\bf y})$ (\ref{cost_function}):
\beq \label{fcost}
{\bf x} = \underset{{\bf x}}{\argmin} f({\bf x},{\bf y})
,\eeq
where ${\bf x}$ are the parameters and ${\bf y}$ are the data.
At a local minimum, the gradient of the cost ${\nabla}f({\bf x},{\bf y})={\bf 0}$. Therefore, in order to find the local minimum, an essential ingredient is the gradient or the derivative. In stochastic optimization, this gradient is approximated \citep{robbins1951} using a subset of the data $\tilde{\bf y}$, i.e., ${\nabla}f({\bf x},{\bf y})\approx {\nabla}f({\bf x},\tilde{\bf y})$. The point of this approximation is that the cost of computing ${\nabla}f({\bf x},\tilde{\bf y})$ is much less than the cost of computing ${\nabla}f({\bf x},{\bf y})$.

For solving problems similar to (\ref{fcost}), we developed a stochastic version of the limited-memory Broyden Fletcher Goldfarb Shanno \citep[LBFGS;][]{Liu1989,Byrd1995} algorithm using this idea of stochastic approximation \citep{DSW2019,Y2020}. In addition to the reduced computational cost, by using the LBFGS algorithm, we were able to achieve faster (super-linear) convergence compared to popular first-order optimization methods based on gradient descent \citep{Y2020}. We applied the same algorithm for IDG-CAL, the only exception being that instead of using subsets of data, we used subsets of sub-grids (sub-grids are explained in Section \ref{sec:model}) to approximate the gradient (and the cost function). It is also possible to divide data into subsets according to the frequency (when there are multiple frequencies or separate channels), but we leave this optimization for future work. The overall stochastic calibration scheme is shown in Algorithm \ref{alg_opt}. In this algorithm, we used the term mini-batch to denote a subset of data (according to sub-grids or channels). An epoch is defined as a sequence of mini-batches that completes the full set of data being calibrated. With each mini-batch of data, a small number of iterations of the LBFGS algorithm is performed. The total amount of data is divided into the number of given mini-batches, and therefore, in each iteration, the amount of data being used is smaller than the total amount of data being calibrated. However, since one epoch includes the full number of mini-batches, during one epoch, the total amount of data is indeed used to obtain the calibration solutions.
\begin{algorithm}
\caption{Stochastic optimization}\label{alg_opt}
\begin{algorithmic}[1]
\REQUIRE Epochs ($E$), mini-batches ($M$), number of iterations ($K$)
\STATE epoch = 0
\FOR{epoch < $E$}
\STATE minibatch = 0
\FOR{minibatch < $M$}
\STATE iteration = 0
\FOR{iteration < $K$}
\STATE Update parameters by minimizing (\ref{fcost}) using the selected minibatch
\STATE iteration++
\ENDFOR
\STATE minibatch++
\ENDFOR
\STATE epoch++
\ENDFOR
\STATE Output solution
\end{algorithmic}
\end{algorithm}

In Table \ref{config}, we show the timing and final error results of a typical calibration with varying mini-batch and epoch sizes. This example uses data with 30 channels that are divided into blocks of ten (three blocks), and the sub-grid size is $48 \times 48$, making the total numbers of sub-grids equal to $2304$ (image size is 8k$\times$8k pixels).

\begin{table*}[htbp]
\begin{minipage}{0.98\linewidth}
\caption{Variation of computational time and final error with various configuration parameters.\label{config}}
\begin{center}
\begin{tabular}{llllll}
\thead{Iterations} & \thead{Epochs} & \thead{Minibatches} & \thead{Total time} & \thead{Time per iteration} & \thead{Normalized error}\\\hline
48 & 1 & 1 & 4590 & 95 & 1.0000\\
8 & 3 & 2 &  1614 & 26 &  1.0156\\
5 & 3  & 3 & 1028 & 17.5 & 1.0251\\
4 & 3  & 4 & 892 &  14.5 & 1.0250\\
3 & 3  & 5 & 668 & 12.5 & 1.0356\\
3 & 3 &  6 & 663 & 9.5 & 1.0446\\
3 & 3  & 7 & 617 & 7.5 & 1.0393\\
\end{tabular}
\end{center}
\end{minipage}
\end{table*}
\begin{figure}[htbp]
\begin{minipage}{0.98\linewidth}
\begin{center}
\centering
\centerline{\includegraphics[width=0.9\textwidth]{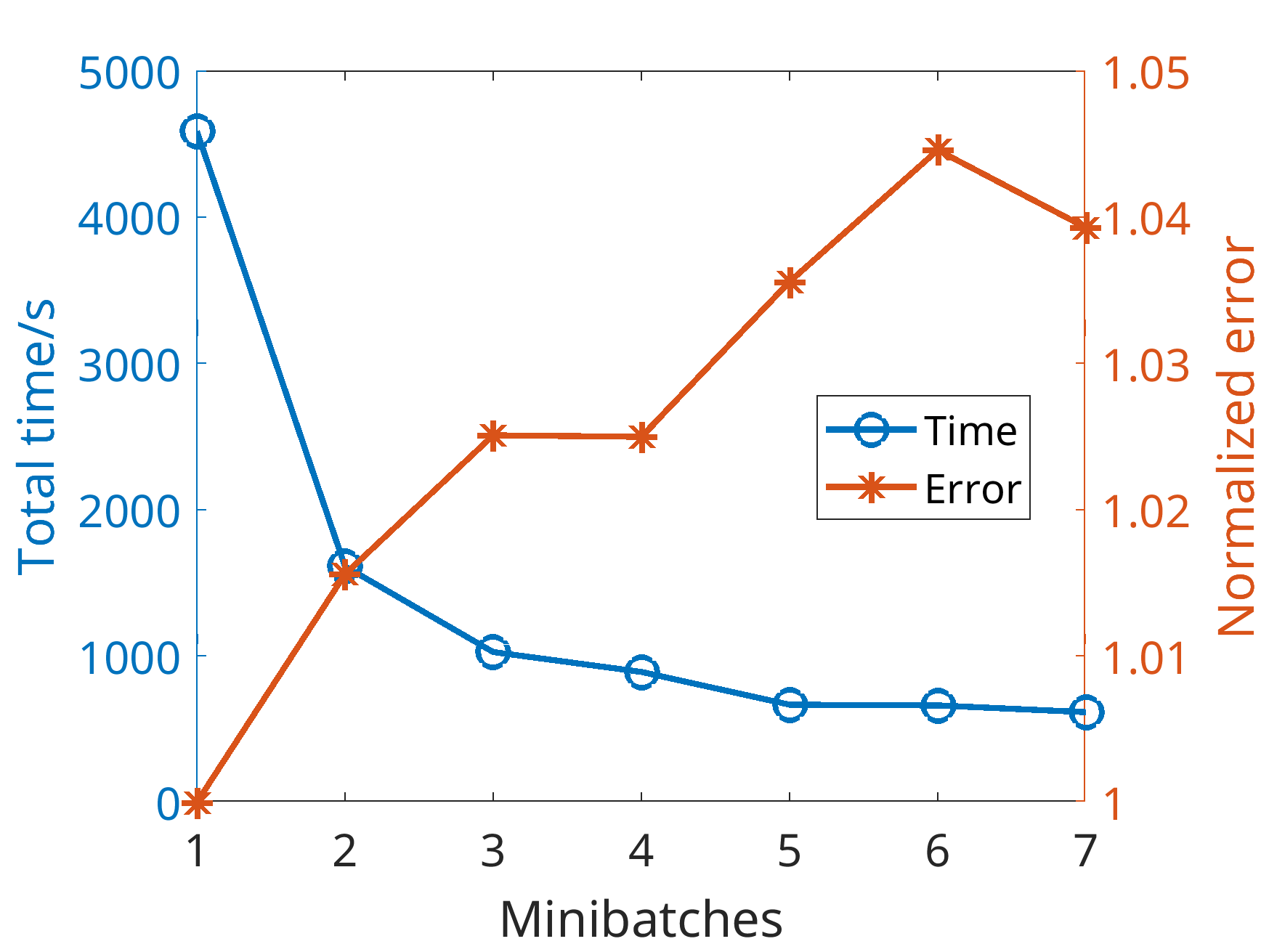}}
\end{center}
\caption{Total time and final error variation with the number of mini-batches. A large number of mini-batches implies a small mini-batch size, which, hence, reduces the computational cost. The error is normalized by the error reached with the full-batch mode of solving.\label{timings}}
\end{minipage}
\end{figure}

In Fig. \ref{timings} we illustrate the variation of the total computation time and the final error with the mini-batch size (values extracted from Table \ref{timings}). We note two distinct behaviors in Fig. \ref{timings}. First, we should remember that the amount of data (or sub-grids) used by a single minibatch is inversely proportional to the number of mini-batches. In Fig. \ref{timings}, the x-axis shows the number of mini-batches increasing from one to seven. When the number of mini-batches is equal to  one, we have the original full-batch mode of optimization. In full-batch mode, all data are used in the evaluation of (\ref{fcost}) per iteration, resulting in the highest computational cost (or computational time). In contrast, as the number of mini-batches increases, the amount of data (or sub-grids) that are being used to evaluate (\ref{fcost}) and its derivative per iteration is reduced, hence reducing the computational time.

An unfavorable consequence of using mini-batches to solve (\ref{fcost}) is the increase in the final error (or the final residual). However, as seen in Fig. \ref{timings}, this increase is very low (a few percent) compared to the error attained by the full-batch mode of calibration. We reached a compromise between the computational cost and the error by using a mini-batch size of $3$ or $4$, which is what we used in our tests discussed in Section \ref{sec:results}.

\section{Results\label{sec:results}}
To illustrate and compare the performance of the proposed calibration scheme, we used a LOFAR high-band array (HBA) observation that is typical of standard LOFAR HBA observations, both in terms of observational setup and of radio frequency interference and ionospheric conditions. The data were observed in 2017 as part of the LC9 019 program. The target (NGC 6338) was observed by LOFAR for eight~hours using 48~MHz of bandwidth in the HBA-low band (centered at $\approx 144$~MHz). The target observation was preceded by a ten-minute calibrator observation of 3C380. The calibrator and target data were obtained from the LOFAR long-term archive (stored with SAS IDs 632471 and 632475 for the calibrator and target data, respectively) and processed with version 5.0 of the LOFAR initial calibration (LINC) pipeline to remove instrumental effects and perform a basic direction-independent calibration on the target field. The data were then averaged to 8~s per time slot and 0.1 MHz per frequency channel, as is typical for HBA data  \citep[e.g.,][]{LoTSS}. International LOFAR stations were not included in the processing. For the tests described below, a subset of 125 frequency channels ($\approx 12$~MHz), centered at $\approx 144$~MHz, was used.

\begin{figure*}[htbp]
\begin{minipage}{1.00\linewidth}
\begin{center}
\centering
\centerline{\includegraphics[width=1.00\textwidth]{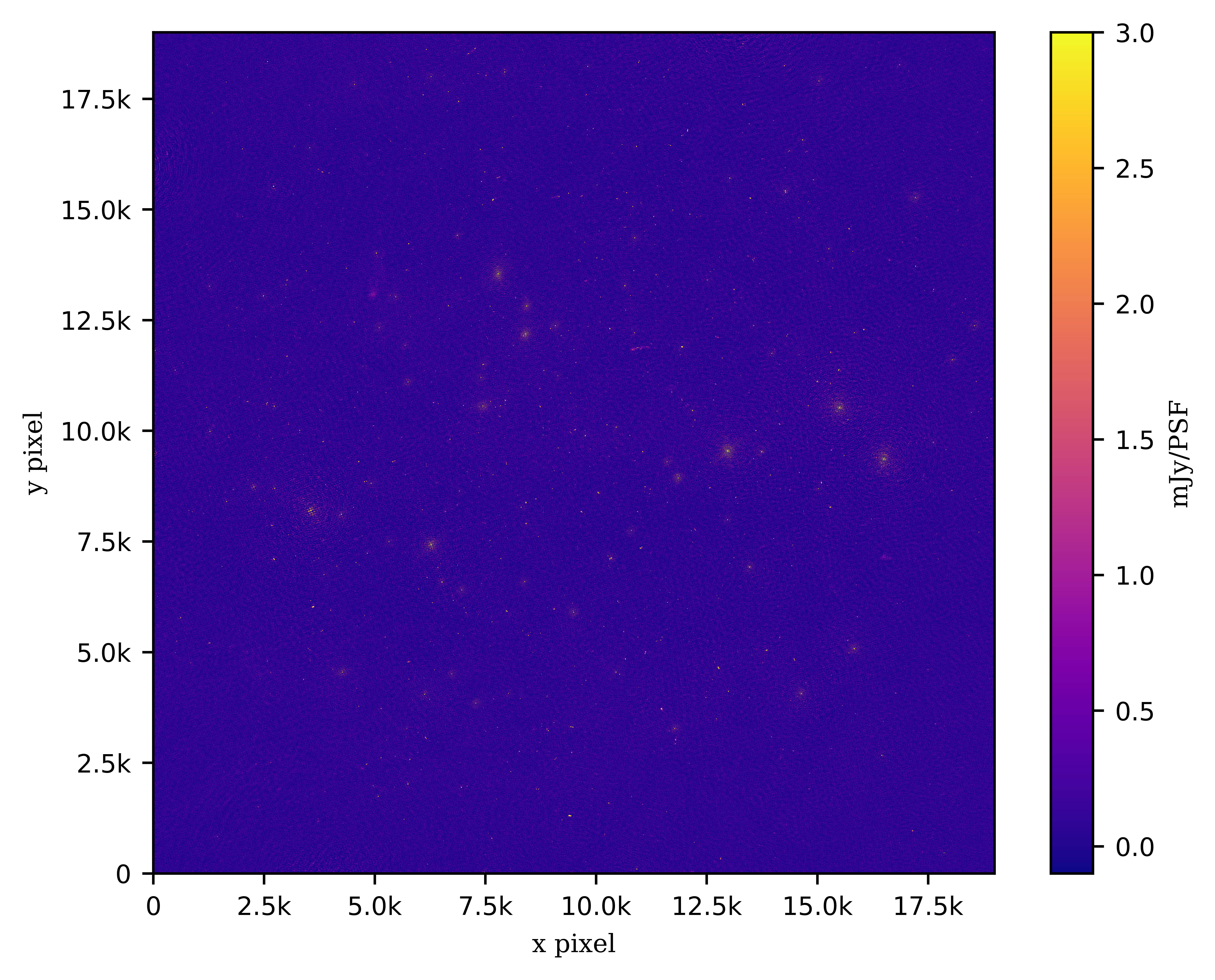}}
\end{center}
\caption{Full field-of-view image produced by standard facet-based calibration after four major cycles. The image size is 19k$\times$19k pixels of $1.25^{\prime\prime}\times 1.25^{\prime\prime}$ covering a field of view of about 7$\times$7 square degrees. The color-scale is linear within the range of [-0.1, 3] mJy/PSF.\label{rapthor_full_fov}}
\end{minipage}
\end{figure*}

In Fig. \ref{rapthor_full_fov}, we show the image produced by facet-based imaging as implemented in the Rapthor pipeline. This image was made after running four major cycles where each one involves a sky-model update, direction-dependent calibration (using DDECal), and facet-based imaging. There are 34 facets covering the field of view of about 7$\times$7 square degrees.

\begin{figure*}[htbp]
\begin{minipage}{1.00\linewidth}
\begin{center}
\centering
\centerline{\includegraphics[width=1.00\textwidth]{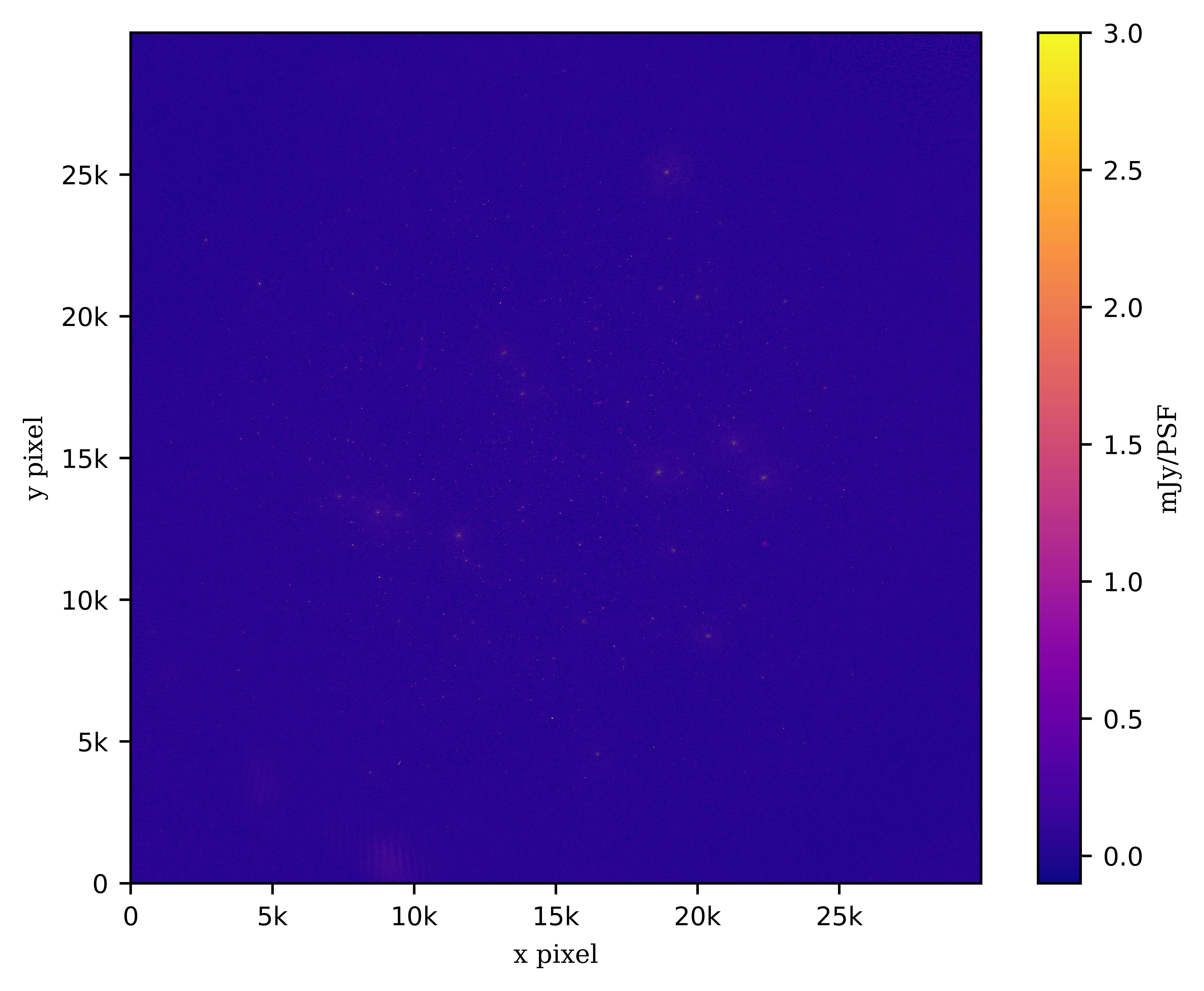}}
\end{center}
  \caption{Full field-of-view image produced by IDG-CAL after three major cycles. The image size is 30k$\times$30k pixels of $1.2^{\prime\prime}\times 1.2^{\prime\prime}$ covering a field of view of about 10$\times$10 square degrees. The color-scale is linear within the range of [-0.1, 3] mJy/PSF (the same as in Fig. \ref{rapthor_full_fov}).\label{idgcal_full_fov}}
\end{minipage}
\end{figure*}

For comparison, in Fig. \ref{idgcal_full_fov}, we show the result obtained by IDG-CAL after running only three major cycles (instead of four major cycles) using the same data. Since we were not restricted by the number of facets, we were able to image a larger 10$\times$10 square degrees in Fig. \ref{idgcal_full_fov}. For a facet-based approach, imaging a larger field of view would require increasing the number of facets used, which would thus increase the computational cost and make the solutions less constrained (more degrees of freedom). In order to run IDG-CAL, we used 32$\times$32 sub-grids with fourth-order polynomials to model the amplitude and phase in (\ref{Apoly}). The number of degrees of freedom for both facet-based calibration and for IDG-CAL are almost the same in this comparison. Note that the artifacts at the bottom left edge of Fig. \ref{idgcal_full_fov} are mainly due to the limitations of the chosen basis functions and the tapering parameters. Such artifacts can be minimized by fine-tuning the aforementioned parameters.

\begin{figure*}[htbp]
\begin{minipage}{0.98\linewidth}
\begin{center}
\centering
\centerline{\includegraphics[width=0.99\textwidth]{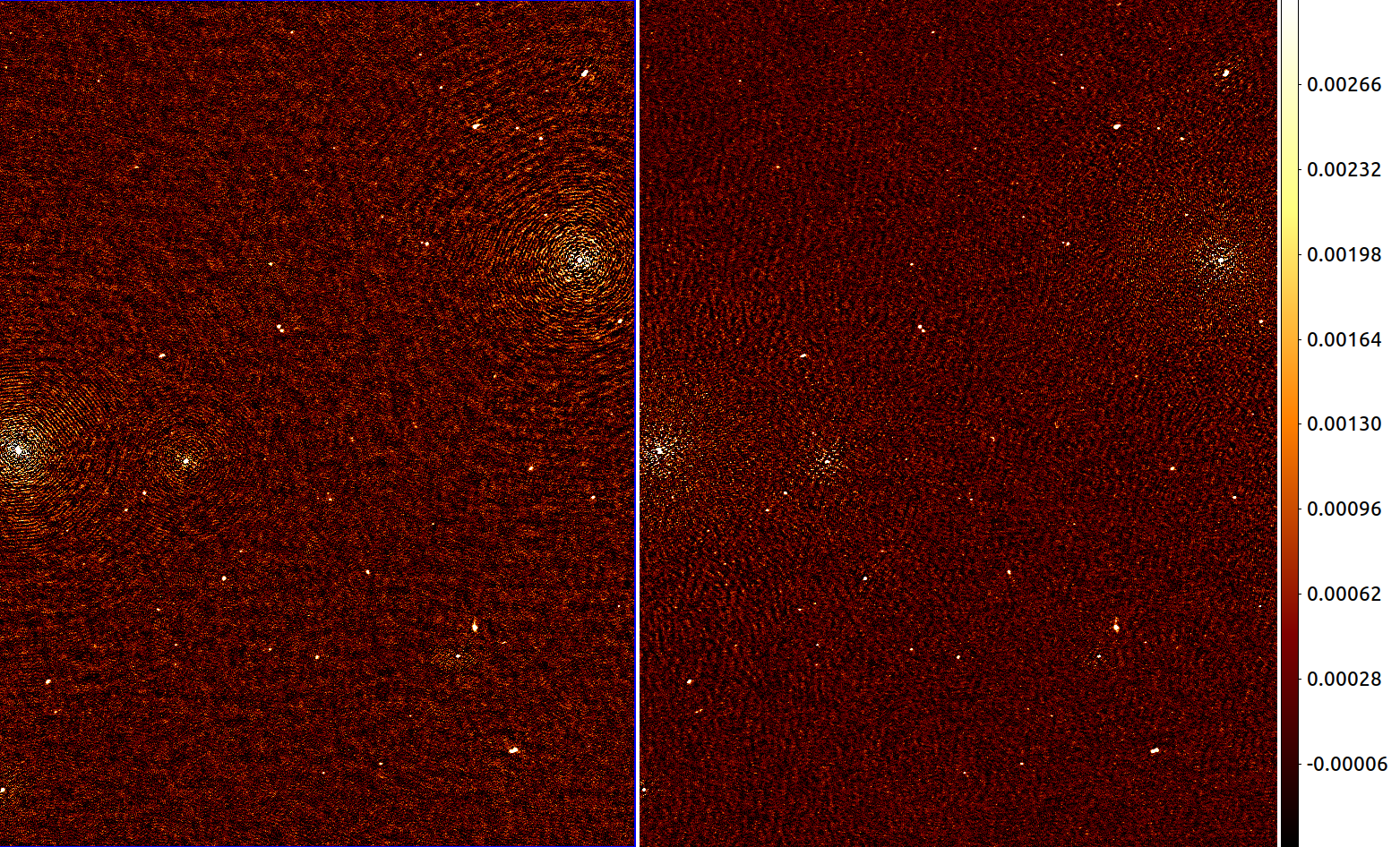}}
\end{center}
\caption{Closeup view of a small part (1$\times$1.3 square degrees) of the images from facet calibration using Rapthor (left) and that from IDG-CAL (right). The color-scale is in Jy/PSF and is shown on the right hand side and is the same for both panels.\label{rapthor_idgcal}}
\end{minipage}
\end{figure*}

In order to make a more qualitative comparison, we show enlarged parts of both Fig. \ref{rapthor_full_fov} and Fig. \ref{idgcal_full_fov} covering the same region in Fig. \ref{rapthor_idgcal}. Both panels in Fig. \ref{rapthor_idgcal} have the same color-scale. The left panel in Fig. \ref{rapthor_idgcal} was made using facet-based calibration, while the right panel in Fig. \ref{rapthor_idgcal} is made using IDG-CAL.

We clearly see that the IDG-CAL image in the right panel of Fig. \ref{rapthor_idgcal} has fewer artifacts around the bright sources. The background noise level is also about 30\% lower in the IDG-CAL image for this reason. In Section \ref{sec:discussion}, we discuss the differences between two approaches in more detail.

\section{Discussion\label{sec:discussion}}
In this section, we discuss the unique characteristics of IDG-CAL in comparison to traditional facet-based calibration as well as its limitations and potential future improvements.
The scalability of calibration algorithms is critical for next-generation LOFAR processing. IDG-CAL offers a more scalable approach than the traditional facet-based DDECal (e.g., \cite{ST2015}) by representing direction-dependent effects as a compact image-domain basis rather than as many independent directional solutions.

In DDECal, calibration is done per facet, with each facet requiring a separate model prediction of visibilities. This causes data duplication, high memory usage, and limited computational intensity. Runtime grows with the number of facets, which increases steeply for wide-field, high-resolution imaging. In the context of the LOFAR enhanced network for sharp surveys (LENSS), which expands LOFAR’s instantaneous field of view by $4{\times}$, this facet approach becomes impractical as the number of facets and the replicated model data grow rapidly.

In contrast, IDG-CAL models the direction dependent effects using image-domain polynomials defined on sub-grids. The computational cost scales roughly linearly with the number of polynomial parameters, not with the number of directions. The method is computation-bound rather than memory-bound; dense kernels dominated by trigonometric and complex arithmetic define time-to-solution methods. These kernels map onto graphics processing units (GPUs) efficiently; their special function units handle trigonometric operations effectively. Modern GPUs deliver a large energy-efficiency advantage for this workload (e.g., ${\sim}223$ GFLOPs/W for a Blackwell GPU versus ${\sim}5.6$ GFLOPs/W for a Haswell CPU).

As seen in Section \ref{sec:results}, IDG-CAL can reach calibration quality comparable to a full facet-based calibration with about three major cycles. The present CPU implementation of IDG-CAL is only slower than DDECal by about a factor of four, but the algorithm’s parallel structure makes it well-suited to GPU acceleration. GPU speedups of about $10{\times}$ over an optimized CPU implementation have been demonstrated for IDG \citep{Veenboer2017}. We therefore expect similar speed improvements of IDG-CAL on accelerator hardware, making it computationally viable.

For LENSS, which increases processed data volume by $4{\times}$ and targets images up to 100k $\times$ 100k pixels, both methods scale with data volume. IDG-CAL preserves predictable computational scaling and modest memory use by increasing sub-grid size and polynomial order. DDECal would require many more facets, causing large increases in model replication, memory footprint, and I/O.

Summarizing the computational advantages, IDG-CAL scales linearly with model complexity, maps efficiently to GPUs, and avoids the memory and I/O bottlenecks of facet-based calibration. These properties align with LENSS goals for sustainable, high-resolution, wide-field calibration.

The expectation is that a parameterized continuous screen would match physical reality better than the piece-wise constant correction with facets. Evidence that this is indeed the case is given by the background noise in the IDG-CAL self-calibration run compared to a facet-based run in Fig. \ref{rapthor_idgcal}.
The Rapthor run uses 24, 31, 33, and 34 facets for the four consecutive self-calibration cycles. In contrast, the IDG-CAL run uses second-, third-, and fourth-order polynomials; i.e., 6, 10, and 15 degrees of freedom in three self-calibration cycles.
As seen in Fig. \ref{rapthor_idgcal}, in the third major cycle, with a fourth-order polynomial with 15 degrees of freedom covering an area of $10\times10$ square degrees, we obtain a lower background noise (\qty{345}{\micro\Jansky}) than the fourth cycle with 34 facets covering an area of $7\times7$ square degrees (\qty{445}{\micro\Jansky}) with traditional facet-based calibration. Summarizing the quality advantage, we find that the physical plausibility of the A-term-based model for the direction-dependent systematic errors provides a model that consumes fewer degrees of freedom, as well as a model that preserves the continuity and, for the most part, the differentiability throughout the full field of view.

We present the first results based on IDG-CAL using real data in this work. We have many future improvements planned that we briefly list below.

\begin{itemize}
  \item GPU acceleration. As discussed above, the IDG-CAL approach is amenable to GPU acceleration. We will explore this together with the use of reduced-precision floating-point calculations when feasible.
  \item Frequency-dependent sky model. So far, we have used model images that are made for narrow frequency bands. We will explore creating frequency-dependent model images where the pixels have spectral information as well as the intensity. Alternatively, we will explore the use of a stack of model images for different frequencies that can be interpolated for the desired frequency.
  \item Better basis functions. We will explore the use of more refined basis functions for modeling the A-term.
  \item Spectral regularization. We will explore the use of spectral regularization \citep{DCAL} to further improve the quality of calibration by constraining the frequency variation of the parameters.
\end{itemize}

The only limitation of IDG-CAL is its intimate link to an image. In facet-based calibration, the sky model can have discrete components and can work with sources that are far apart without creating an image covering all sources that are present. A way to overcome this limitation is to subtract the strong outlier sources from the data (that are not of scientific interest in order to make images) before running IDG-CAL.

\section{Conclusions\label{sec:conclusions}}
We introduce IDG-CAL, an alternative method for direction-dependent calibration of radio interferometric data using a physically plausible A-term based model. In comparison to traditional facet-based calibration, we see improved image quality as well as the possibility for computational efficiency especially when imaging large fields of view. Further improvements of IDG-CAL are being planned and will improve both the quality of the calibration solutions as well as make the computations run even more efficiently. 

\begin{acknowledgements}
We thank the reviewer and editor for their comments. This work has been supported by FuSE: Fundamental Sciences e-Infrastructure, Dutch Research Council (NWO) Roadmap Project 184.035.004.
 LOFAR, the Low Frequency Array designed and constructed by ASTRON, has facilities in several countries, owned by various parties (each with their own funding sources), and collectively operated by the International LOFAR Telescope (ILT) foundation under a joint scientific policy.

  The software used in this paper are: \textsc{IDG} (\url{https://git.astron.nl/RD/idg}),  \textsc{DP3} (\url{https://git.astron.nl/RD/DP3}) \citep{DP3}, \textsc{Rapthor} (\url{https://git.astron.nl/RD/rapthor}), \textsc{SAGECal} (\url{https://sagecal.sourceforge.net/}), and  WSClean \citep{WSClean}.
\end{acknowledgements}
\bibliographystyle{aa}
\bibliography{references}

@ARTICLE{LOFAR,
   author = {{van Haarlem}, M.~P. and {Wise}, M.~W. and {Gunst}, A.~W. and  others},
    title = "{LOFAR: The LOw-Frequency ARray}",
  journal = {\aap},
archivePrefix = "arXiv",
   eprint = {1305.3550},
 primaryClass = "astro-ph.IM",
     year = 2013,
    month = aug,
   volume = 556,
      eid = {A2},
    pages = {A2},
      doi = {10.1051/0004-6361/201220873},
}

@ARTICLE{MWA,
   author = {{Tingay}, S.~J. and {Goeke}, R. and {Bowman}, J.~D. and {Emrich}, D. and others},
    title = "{The Murchison Widefield Array: The Square Kilometre Array Precursor at Low Radio Frequencies}",
  journal = {\pasa},
archivePrefix = "arXiv",
   eprint = {1206.6945},
 primaryClass = "astro-ph.IM",
     year = 2013,
    month = jan,
   volume = 30,
      eid = {e007},
    pages = {e007},
      doi = {10.1017/pasa.2012.007},
}

@ARTICLE{IDG,
       author = {{van der Tol}, Sebastiaan and {Veenboer}, Bram and {Offringa}, Andr{\'e} R.},
        title = "{Image Domain Gridding: a fast method for convolutional resampling of visibilities}",
      journal = {\aap},
         year = 2018,
        month = aug,
       volume = {616},
          eid = {A27},
        pages = {A27},
          doi = {10.1051/0004-6361/201832858},
archivePrefix = {arXiv},
       eprint = {1909.07226},
 primaryClass = {astro-ph.IM},
}

@article{Kaz2,
 author={{Kazemi}, S. and {Yatawatta}, S. and {Zaroubi}, S. and {Labropoluos}, P. and  {de Bruyn}, A.G. and {Koopmans}, L. and {Noordam}, J.},
 title={Radio interferometric calibration using the {SAGE} algorithm},
 journal = {\mnras},
 volume={414},
 number={2},
 pages={1656--1666},
 month=Jun,
 year={2011}
}

@article{DCAL,
author = {{Yatawatta}, S.}, 
title = {Distributed radio interferometric calibration},
volume = {449}, 
number = {4}, 
pages = {4506-4514}, 
year = {2015}, 
doi = {10.1093/mnras/stv596}, 
journal = {\mnras},
}

@ARTICLE{Tasse2018,
       author = {{Tasse}, C. and {Hugo}, B. and {Mirmont}, M. and {Smirnov}, O. and {Atemkeng}, M. and {Bester}, L. and {Hardcastle}, M.~J. and {Lakhoo}, R. and {Perkins}, S. and {Shimwell}, T.},
        title = "{Faceting for direction-dependent spectral deconvolution}",
      journal = {\aap},
         year = 2018,
        month = apr,
       volume = {611},
          eid = {A87},
        pages = {A87},
          doi = {10.1051/0004-6361/201731474},
archivePrefix = {arXiv},
       eprint = {1712.02078},
 primaryClass = {astro-ph.IM},
       adsurl = {https://ui.adsabs.harvard.edu/abs/2018A&A...611A..87T}
}

@ARTICLE{vanWeeren2016,
       author = {{van Weeren}, R.~J. and {Williams}, W.~L. and {Hardcastle}, M.~J. and {Shimwell}, T.~W. and {Rafferty}, D.~A. and {Sabater}, J. and {Heald}, G. and {Sridhar}, S.~S. and {Dijkema}, T.~J. and {Brunetti}, G. and {Br{\"u}ggen}, M. and {Andrade-Santos}, F. and {Ogrean}, G.~A. and {R{\"o}ttgering}, H.~J.~A. and {Dawson}, W.~A. and {Forman}, W.~R. and {de Gasperin}, F. and {Jones}, C. and {Miley}, G.~K. and {Rudnick}, L. and {Sarazin}, C.~L. and {Bonafede}, A. and {Best}, P.~N. and {B{\^\i}rzan}, L. and {Cassano}, R. and {Chy{\.z}y}, K.~T. and {Croston}, J.~H. and {Ensslin}, T. and {Ferrari}, C. and {Hoeft}, M. and {Horellou}, C. and {Jarvis}, M.~J. and {Kraft}, R.~P. and {Mevius}, M. and {Intema}, H.~T. and {Murray}, S.~S. and {Orr{\'u}}, E. and {Pizzo}, R. and {Simionescu}, A. and {Stroe}, A. and {van der Tol}, S. and {White}, G.~J.},
        title = "{LOFAR Facet Calibration}",
      journal = {\apjs},
         year = 2016,
        month = mar,
       volume = {223},
       number = {1},
          eid = {2},
        pages = {2},
          doi = {10.3847/0067-0049/223/1/2},
archivePrefix = {arXiv},
       eprint = {1601.05422},
 primaryClass = {astro-ph.IM},
       adsurl = {https://ui.adsabs.harvard.edu/abs/2016ApJS..223....2V}
}

@ARTICLE{WBA,
   author = {{Bhatnagar}, S. and {Rau}, U. and {Golap}, K.},
    title = "{Wide-field wide-band Interferometric Imaging: The WB A-Projection and Hybrid Algorithms}",
  journal = {\apj},
archivePrefix = "arXiv",
   eprint = {1304.4987},
 primaryClass = "astro-ph.IM",
     year = 2013,
    month = jun,
   volume = 770,
      eid = {91},
    pages = {91},
      doi = {10.1088/0004-637X/770/2/91},
   adsurl = {http://adsabs.harvard.edu/abs/2013ApJ...770...91B}
}

@ARTICLE{B2008,
       author = {{Bhatnagar}, S. and {Cornwell}, T.~J. and {Golap}, K. and {Uson}, J.~M.},
        title = "{Correcting direction-dependent gains in the deconvolution of radio interferometric images}",
      journal = {\aap},
         year = 2008,
        month = aug,
       volume = {487},
       number = {1},
        pages = {419-429},
          doi = {10.1051/0004-6361:20079284},
archivePrefix = {arXiv},
       eprint = {0805.0834},
 primaryClass = {astro-ph},
       adsurl = {https://ui.adsabs.harvard.edu/abs/2008A&A...487..419B}
}

@ARTICLE{Aproj,
   author = {{Tasse}, C. and {van der Tol}, S. and {van Zwieten}, J. and 
	{van Diepen}, G. and {Bhatnagar}, S.},
    title = "{Applying full polarization A-Projection to very wide field of view instruments: An imager for LOFAR}",
  journal = {\aap},
archivePrefix = "arXiv",
   eprint = {1212.6178},
 primaryClass = "astro-ph.IM",
     year = 2013,
    month = may,
   volume = 553,
      eid = {A105},
    pages = {A105},
      doi = {10.1051/0004-6361/201220882},
   adsurl = {http://adsabs.harvard.edu/abs/2013A%26A...553A.105T}
}

@ARTICLE{Intema,
   author = {{Intema}, H.~T. and {van der Tol}, S. and {Cotton}, W.~D. and 
	{Cohen}, A.~S. and {van Bemmel}, I.~M. and {R{\"o}ttgering}, H.~J.~A.
	},
    title = "{Ionospheric calibration of low frequency radio interferometric observations using the peeling scheme. I. Method description and first results}",
  journal = {\aap},
archivePrefix = "arXiv",
   eprint = {0904.3975},
 primaryClass = "astro-ph.IM",
     year = 2009,
    month = jul,
   volume = 501,
    pages = {1185-1205},
      doi = {10.1051/0004-6361/200811094}
}

@ARTICLE{Roth2023,
       author = {{Roth}, Jakob and {Arras}, Philipp and {Reinecke}, Martin and {Perley}, Richard A. and {Westermann}, R{\"u}diger and {En{\ss}lin}, Torsten A.},
        title = "{Bayesian radio interferometric imaging with direction-dependent calibration}",
      journal = {\aap},
         year = 2023,
        month = oct,
       volume = {678},
          eid = {A177},
        pages = {A177},
          doi = {10.1051/0004-6361/202346851},
archivePrefix = {arXiv},
       eprint = {2305.05489},
 primaryClass = {astro-ph.IM},
       adsurl = {https://ui.adsabs.harvard.edu/abs/2023A&A...678A.177R}
}

@ARTICLE{Arras2019,
       author = {{Arras}, Philipp and {Frank}, Philipp and {Leike}, Reimar and {Westermann}, R{\"u}diger and {En{\ss}lin}, Torsten A.},
        title = "{Unified radio interferometric calibration and imaging with joint uncertainty quantification}",
      journal = {\aap},
         year = 2019,
        month = jul,
       volume = {627},
          eid = {A134},
        pages = {A134},
          doi = {10.1051/0004-6361/201935555},
archivePrefix = {arXiv},
       eprint = {1903.11169},
 primaryClass = {astro-ph.IM},
}

@ARTICLE{Birdi2020,
       author = {{Birdi}, Jasleen and {Repetti}, Audrey and {Wiaux}, Yves},
        title = "{Polca SARA - full polarization, direction-dependent calibration, and sparse imaging for radio interferometry}",
      journal = {\mnras},
         year = 2020,
        month = mar,
       volume = {492},
       number = {3},
        pages = {3509-3528},
          doi = {10.1093/mnras/stz3555},
archivePrefix = {arXiv},
       eprint = {1904.00663},
 primaryClass = {astro-ph.IM},
       adsurl = {https://ui.adsabs.harvard.edu/abs/2020MNRAS.492.3509B}
}

@article{Sault1996,
	author = {{Sault, R. J.} and {Staveley-Smith, L.} and {Brouw, W. N.}},
	title = {An approach to interferometric mosaicing},
	DOI= "10.1051/aas:1996287",
	url= "https://doi.org/10.1051/aas:1996287",
	journal = {Astron. Astrophys. Suppl. Ser.},
	year = 1996,
	volume = 120,
	number = 2,
	pages = "375-384",
}

@article{robbins1951,
author = "Robbins, Herbert and Monro, Sutton",
doi = "10.1214/aoms/1177729586",
fjournal = "The Annals of Mathematical Statistics",
journal = "Ann. Math. Statist.",
month = "09",
number = "3",
pages = "400--407",
publisher = "The Institute of Mathematical Statistics",
title = "A Stochastic Approximation Method",
url = "https://doi.org/10.1214/aoms/1177729586",
volume = "22",
year = "1951"
}

@INPROCEEDINGS{DSW2019,
author={S. {Yatawatta} and L. {De Clercq} and H. {Spreeuw} and F. {Diblen}},
booktitle={2019 IEEE Data Science Workshop (DSW)},
title={A Stochastic {LBFGS} Algorithm for Radio Interferometric Calibration},
year={2019},
volume={},
number={},
pages={208-212},
doi={10.1109/DSW.2019.8755567},
ISSN={null},
month={June},}

@article{Liu1989,
 author = {Liu, D. C. and Nocedal, J.},
 title = {On the Limited Memory {BFGS} Method for Large Scale Optimization},
 journal = {Math. Program.},
 issue_date = {Dec. 1989},
 volume = {45},
 number = {3},
 month = dec,
 year = {1989},
 issn = {0025-5610},
 pages = {503--528},
 numpages = {26},
 url = {http://dx.doi.org/10.1007/BF01589116},
 doi = {10.1007/BF01589116},
 acmid = {83726},
 publisher = {Springer-Verlag New York, Inc.},
 address = {Secaucus, NJ, USA},
}

@article{Byrd1995,
author = {Byrd, Richard H. and Lu, Peihuang and Nocedal, Jorge and Zhu, Ciyou},
title = {A Limited Memory Algorithm for Bound Constrained Optimization},
journal = {SIAM Journal on Scientific Computing},
volume = {16},
number = {5},
pages = {1190-1208},
year = {1995},
doi = {10.1137/0916069},
}

@article{Y2020,
    author = {Yatawatta, Sarod},
    title = "{Stochastic calibration of radio interferometers}",
    journal = {\mnras},
    volume = {493},
    number = {4},
    pages = {6071-6078},
    year = {2020},
    month = {03},
    issn = {0035-8711},
    doi = {10.1093/mnras/staa648},
    url = {https://doi.org/10.1093/mnras/staa648},
    eprint = {https://academic.oup.com/mnras/article-pdf/493/4/6071/32980363/staa648.pdf},
}

@INPROCEEDINGS{Veenboer2017,
  author={Veenboer, Bram and Petschow, Matthias and Romein, John W.},
  booktitle={2017 IEEE International Parallel and Distributed Processing Symposium (IPDPS)}, 
  title={Image-Domain Gridding on Graphics Processors}, 
  year={2017},
  volume={},
  number={},
  pages={545-554},
  doi={10.1109/IPDPS.2017.68}}

@ARTICLE{ST2015,
       author = {{Smirnov}, O.~M. and {Tasse}, C.},
        title = "{Radio interferometric gain calibration as a complex optimization problem}",
      journal = {\mnras},
         year = 2015,
        month = may,
       volume = {449},
       number = {3},
        pages = {2668-2684},
          doi = {10.1093/mnras/stv418},
}

@ARTICLE{WSClean,
       author = {{Offringa}, A.~R. and {McKinley}, B. and others},
        title = "{WSCLEAN: an implementation of a fast, generic wide-field imager for radio astronomy}",
      journal = {\mnras},
         year = 2014,
        month = oct,
       volume = {444},
       number = {1},
        pages = {606-619},
          doi = {10.1093/mnras/stu1368},
archivePrefix = {arXiv},
       eprint = {1407.1943},
 primaryClass = {astro-ph.IM},
}

@software{DP3,
       author = {{van Diepen}, Ger and {Dijkema}, Tammo Jan and {Offringa}, Andr{\'e}},
        title = "{DPPP: Default Pre-Processing Pipeline}",
 howpublished = {Astrophysics Source Code Library, record ascl:1804.003},
         year = 2018,
        month = apr,
          eid = {ascl:1804.003},
archivePrefix = {ascl},
       eprint = {1804.003},
}

@ARTICLE{Rau2009,
       author = {{Rau}, Urvashi and {Bhatnagar}, Sanjay and {Voronkov}, Maxim A. and {Cornwell}, Tim J.},
        title = "{Advances in Calibration and Imaging Techniques in Radio Interferometry}",
      journal = {IEEE Proceedings},
         year = 2009,
        month = aug,
       volume = {97},
       number = {8},
        pages = {1472-1481},
          doi = {10.1109/JPROC.2009.2014853},
archivePrefix = {arXiv},
       eprint = {0902.0817},
 primaryClass = {astro-ph.IM},
}

@ARTICLE{LoTSS,
       author = {{Shimwell}, T.~W. and {R{\"o}ttgering}, H.~J.~A. and {Best}, P.~N. and {Williams}, W.~L. and {Dijkema}, T.~J. and {de Gasperin}, F. and {Hardcastle}, M.~J. and {Heald}, G.~H. and {Hoang}, D.~N. and {Horneffer}, A. and {Intema}, H. and {Mahony}, E.~K. and {Mandal}, S. and {Mechev}, A.~P. and {Morabito}, L. and {Oonk}, J.~B.~R. and {Rafferty}, D. and {Retana-Montenegro}, E. and {Sabater}, J. and {Tasse}, C. and {van Weeren}, R.~J. and {Br{\"u}ggen}, M. and {Brunetti}, G. and {Chy{\.z}y}, K.~T. and {Conway}, J.~E. and {Haverkorn}, M. and {Jackson}, N. and {Jarvis}, M.~J. and {McKean}, J.~P. and {Miley}, G.~K. and {Morganti}, R. and {White}, G.~J. and {Wise}, M.~W. and {van Bemmel}, I.~M. and {Beck}, R. and {Brienza}, M. and {Bonafede}, A. and {Calistro Rivera}, G. and {Cassano}, R. and {Clarke}, A.~O. and {Cseh}, D. and {Deller}, A. and {Drabent}, A. and {van Driel}, W. and {Engels}, D. and {Falcke}, H. and {Ferrari}, C. and {Fr{\"o}hlich}, S. and {Garrett}, M.~A. and {Harwood}, J.~J. and {Heesen}, V. and {Hoeft}, M. and {Horellou}, C. and {Israel}, F.~P. and {Kapi{\'n}ska}, A.~D. and {Kunert-Bajraszewska}, M. and {McKay}, D.~J. and {Mohan}, N.~R. and {Orr{\'u}}, E. and {Pizzo}, R.~F. and {Prandoni}, I. and {Schwarz}, D.~J. and {Shulevski}, A. and {Sipior}, M. and {Smith}, D.~J.~B. and {Sridhar}, S.~S. and {Steinmetz}, M. and {Stroe}, A. and {Varenius}, E. and {van der Werf}, P.~P. and {Zensus}, J.~A. and {Zwart}, J.~T.~L.},
        title = "{The LOFAR Two-metre Sky Survey. I. Survey description and preliminary data release}",
      journal = {\aap},
         year = 2017,
        month = feb,
       volume = {598},
          eid = {A104},
        pages = {A104},
          doi = {10.1051/0004-6361/201629313},
archivePrefix = {arXiv},
       eprint = {1611.02700},
 primaryClass = {astro-ph.IM},
       adsurl = {https://ui.adsabs.harvard.edu/abs/2017A&A...598A.104S}
}
\end{document}